\documentclass[preprint,showpacs,superscriptaddress,preprintnumbers,amsmath,amssymb]{revtex4}

\usepackage{graphicx}			%
\usepackage[utf8]{inputenc} 		%
\usepackage[unicode=true]{hyperref} 	%
\usepackage{dcolumn}

\begin{document}

\title{Does the fireball found in AuAu collisions at RHIC resemble plasma?}
\author{R.Ya.~Zulkarneev}
\affiliation{Joint Institute for Nuclear Research, Dubna}
\date{\today}

\begin{abstract}
General properties of the hot and dense hadron matter (fireball) discovered in AuAu collisions with $\sqrt{S_{NN}}$ = 130 and 200 GeV at RHIC are compared with these for ideal electromagnetic plasma. None of them was found to be in contrary to those of the plasma typical signs distinguishing its from
other aggregate states of the matter. The author notes that modern experimental data about the fireball properties are limited to make their comprehensive comparison with plasma signatures. The author also points out the directions needed to be studied to answer question whether the hadron matter observed in the RHIC experiments is a color analogue of the electromagnetic plasma or not.
\end{abstract}


\maketitle

\newpage
\section {Introduction}

The search for quark-gluon plasma (QGP) has been carried out for about almost three decades. The main features of this hypothetic state of the hadron
matter according to \cite{1}-\cite{10} are as follows:

\begin{itemize}
\item deconfinement of quarks;
\item quarks and gluons are deconfined in the volume scale of two nuclei collision but not in a single nucleon bag;
\item characteristic density of energy in the collision volume is at the order of several GeV/fm$^{3}$ 
\footnote{A definite energy density must be created in the volume of the collisions according to \cite{10} to achieve 
the state where the color charges can turn out to be separated.}
\end{itemize}
There is no experimental confirmation to any of the above except the latter. The same is true concerning the list of features
 obtained by the LQCD, quark and other microscopic models which to define QGP \cite{1},\cite{2}\cite{3}\cite{4}\cite{5}\cite{6},\cite{7},\cite{8},\cite{9}. 
 The discussion of these features is given below.

The author of this work has considered the general properties of the hot and dense hadron matter (fireball) found in AuAu collisions 
with $\sqrt{S_{NN}}$ = 130 and 200 GeV at RHIC to figure out how they correspond to the characteristic properties of the ideal electromagnetic 
plasma (further on ‘‘plasma’’). The results of the performed analysis have shown that some properties of the fireball (see Table \ref{tab:tb_01}) resemble
the properties of the plasma. The other properties do not contradict this conclusion but need further clarification. Thus, the existing experimental
information on the fireball produced at RHIC is not enough at present to carry out its total completed comparison with the plasma. The work has 
shown that to reach completeness in future comparisons, it is necessary to have the following: 
\begin{itemize}
    \item[a)] more precise experimental data to study the screening effects of the color charges in the fireball medium; 
    \item[b)] direct experimental confirmation of quasi-particle excitations in the fireball medium; 
    \item[c)] direct experimental estimations of the fireball temperature, for example, by measuring the thermal photon spectrum 
in the hot phase of the fireball development. 
\end{itemize}
The experiments mentioned above can give a concluding answer to the question: Is the hadron 
matter observed at RHIC a color analogue of the plasma?

\section{Main properties of the electromagnetic plasma. Their comparison with features of the hadron matter produced in heavy nuclei collisions at RHIC and SPS CERN.}

Let us review the main properties of the electromagnetic plasma. According to \cite{11},\cite{12} plasma is a state of substance having a set 
of the following features:
\begin{enumerate}
    \item Quasineutrality ( or neutrality) on the electric charge of the system in general; \label{p1}
    \item Presence of separated charges ( the charge dissociation of the atom); \label{p2}
    \item Collective character of particle interactions of the system; \label{p3}
    \item The presence of the wave motions in the system beside its chaoticity (quasi-particle degrees of freedom); \label{p4}
    \item Charge screening in the plasma medium. \label{p5}
\end{enumerate}

Besides, the plasma (not in the outside field) is characterized by the following microscopic scales reflecting its properties 
mentioned above and some others: ionization energy of the plasma atoms, Langmuir oscillation frequency and the radius of Debay charge screening - $r_{D}$.

All varieties of plasma in classical physics are known to have these features in one or another extent. These properties can be also considered as 
plasma macroscopic signatures to identify the states of the matter. Does the hadron matter produced in heavy nuclei collisions at ultra relativistic 
RHIC energies, possess similar properties? The question has not been discussed in the literature in this configuration yet. To get an answer, 
let us consider some experimental facts and conclusions from their analysis obtained earlier. Our consideration will be qualitative. While comparing 
we suppose that the analogue of the plasma electric charge in the hadron matter is, naturally, a color charge.

Since the nucleon is colorless and there are no free quarks in the nature \cite{14} (no free color charges then), condition \ref{p1} will be also 
preserved in the case of the hadron matter, this time as its ‘‘color neutrality’’. So, property \ref{p2} will reflect separation of the color charges 
in the fireball, i.e. deconfinement of quarks in the fireball medium.

The character of the third plasma property - collective interactions of its particles, is determined by the Coulomb forces between the charges in
the plasma which reduce slower with the distance than Van der Vaals forces acting between neutral atoms. It is not evident that similar to the 
electron plasma field there will be the collective forces field established between the fireball constituents. But it turns out that exactly this
collective interaction takes place between the light quarks in the initial hot phase of the fireball evolution at RHIC energies. This was shown by
the hydrodynamic analysis \cite{15} of the elliptic flows of light mesons and baryons in AuAu collisions with $\sqrt{S_{NN}}$ = 200 GeV. Recently an
analogous type of interactions has been found to exist also between the heavy and light quarks in the fireball matter at RHIC. This conclusion goes 
from the analysis of the energy spectra and values of the elliptic flows of fast electrons of the heavy mesons produced from $b-$,$c-$ and other 
quarks \cite{16},\cite{17}. Thus the collective fireball properties are substantiated by serious experimental data.

The other plasma property is related with charge density fluctuations of the plasma in the time. They reveal themselves as the quasi-particles motion
in the plasma matter \cite{13}. In the case of the fireball the appearance of this motion form significantly depends on the color dielectric properties
of the fireball medium. The medium effects open one more area to compare the behavior of these two forms of the matter. So, it has been figured out 
that to take the oscillation degrees of freedom into account (for example, in the form of gluon plasmons \cite{18} in the quark-gluon medium) is 
important while calculating the radiation losses of the parton mini-jet energy when they are moving through this medium. 

First indications on this 
effect to exist in the fireball medium produced in the central collisions at RHIC, were obtained in the work \cite{19}. For the first time it allowed 
one to solve successfully the known problem of strong suppression of the $\pi^{0}$ meson yields in AuAu collisions at 130 GeV. The author has used the
so called quasi-particle model of QGP where the quark-particle excitations and gluons with transverse polarization are described as massive 
quasi-particles with parameters depending on the medium properties \cite{20}. The model is characterized by big energy losses of the gluon and quark 
jets in the QGP medium. This peculiarity gives an opportunity to hope that it will promote a solution of another analogous puzzle - suppression of 
the charmed and other heavier meson states produced in the collisions at RHIC \cite{21},\cite{16}. There is no satisfying explanation of this problem.
All the suppression mechanisms offered earlier (also including the plasmon degrees of freedom in the gluon component of QGP) enable us to obtain only 
the qualitative agreement of the experimental data \cite{22}. In principle, it can be improved if to take two-particle resonances \cite{23} or three-particle 
heavy quark collisions in the medium \cite{17}. But so far the both mechanisms remain only as indications of the opportunity to excite the simplest 
quasi-particle states in the quark-gluon fireball medium. It is needed to study further this important problem to understand whether the heated fireball 
medium (at RHIC) has quasi-particle degrees of freedom.

To obtain direct but not subordinate experimental signals on the existence of these degrees of freedom, would be extremely important in studying the
fireball properties. One of these experiments of this type should carry out collisions of partially ionized atoms of heavy nuclei at RHIC and LHC. 
In this case the electrons of the atom shells of ions will be like an electromagnetic probe to be used, in principle, for direct study of the 
quasiparticle excitations of the produced fireball medium.

Finally about property \ref{p5}. The ability of screening the charges leads to the appearance of the characteristic scale $r_{D}$ in the plasma 
medium which describes the plasma field at distances $r < r_{D}$. As it is seen from the LQGP calculations \cite{24}, the analogous scale appears 
also in the color-charged matter. In the consequence of the above one can expect significant suppression of the production of some atom-similar 
systems which are rather prolonged, for example, $\chi$ and $\psi\prime$ charmonium states. These effects are traditionally called ‘‘melting’’ states. 
The first experiments to observe these melting $J/\psi$, $\psi\prime$, $\chi$ charmonium states on the SPS beams at CERN ($\sqrt{S_{NN}}$ = 17 GeV) 
\cite{21} and later at RHIC ($\sqrt{S_{NN}}$ = 130 GeV) \cite{25},\cite{26} have confirmed the general character of the expectations. But a 
theoretical interpretation of the experimental data still remains to be not clear: not all principally important pecularities of the yield 
of $J/\psi$, $\psi\prime$, $\chi$ meson states have been observed so far \cite{27}. The measurement precision needs to be improved. The study of 
the problem is going on.

In conclusion we have to emphasize that a specific feature of the substance state in the form of plasma is the collective character of its 
constituents interaction and excitations of its medium. Now none of the experimentally known properties of the fireball produced in the collisions 
of heavy nuclei at $\sqrt{S_{NN}}$ = 130 and 200 GeV, contradicts this plasma property. On the contrary, the experimental data have shown that 
some part of them reveals undoubtedly properties similar to \ref{p1}, \ref{p3} and, probably, \ref{p4}. In other cases (properties \ref{p4} and \ref{p5}) 
this similarity in the properties does not contradict to the experiment \cite{17},\cite{19},\cite{21},\cite{25},\cite{27} but it does not result 
from them with enough argumentation either. So that is why it needs to be reliably confirmed either. The summary of the properties of the plasma 
and fireball produced in the collisions of heavy nuclei at collider RHIC and SPS, is shown in Table \ref{tab:tb_01}.
	
\section {Comparison of QGP and electron plasma signatures}

For many years the QGP features obtained on QCD calculations \cite{1}-\cite{10} (see Table \ref{tab:tb_02}.) were the only characteristics to 
identify QGP in the experiments at RHIC and SPS. It is a natural wish to clarify how these properties agree with the plasma signatures \ref{p1}-\ref{p5}. 
The results of comparison are given below. It is easy to notice that the features shown in the first three columns of Table \ref{tab:tb_02} with the sign 
‘‘+’’, are adequate to the classical notion of plasma and prevail in most of the versions of the QGP definition. It is worth emphasizing that the 
Debay radius $r_{D}$ should have been taken as the third feature which jointly with the condition \ref{p1} defines the linear sizes of the plasma.

Independence of the quark motion in QGP (feature 4 in Table \ref{tab:tb_02}) cannot go together with the main property of the electron plasma - the 
collective interaction of its constituents. Thermalization (as feature 4 in Table \ref{tab:tb_02}) is not obligatory a typical attribute of the plasma. 
The following question is still under discussion by theorists: Is the restoration of the chiral symmetry a typical feature of QGP?

And finally about the smallness of the quark mass (feature 6 in Table \ref{tab:tb_02}). The mass of the charge carrier in the electron plasma is not its 
characteristic feature either. You see, the experimental data at RHIC \cite{28} have shown that the masses of the fireball constituents are close to the 
constituent but not the	current values of the quark masses. Thus, not all QGP features offered to identify it in experiments can go with the notion 
of plasma in the classical physics. The list of QGP features \cite{1,9} is lacking in its important collective properties related, in particular, with its 
oscillatory, i.e. quasiparticle degrees of freedom.
		
\section {Conclusion}

So, the main properties of the hot and dense hadron matter observed at RHIC turned out to resemble more the usual electron plasma than the image of 
QGP which appeared from the simplest quark descriptions. The fireball medium did not behave like an ideal gas of particles. On the contrary, its 
constituents turned out to be strongly bound. They interact with hadrons by $\sim 50$ times stronger than the casual hadron matter and turned out 
for them to be a rather opaque medium. Only this has motivated the title of this medium as ‘‘plasma’’ with strongly interacting quarks and gluons 
(sQGP) \cite{9},\cite{29}. In the meanwhile, to identify this state of the matter produced at RHIC as the state of the plasma, it is necessary to 
obtain positive answers to the questions which still remain unclear:

\begin{itemize}
 \item is the medium of the produced fireball able to excite wave degrees of freedom in it?
 \item can the fireball medium screen the color charges?
 \item what is the nature of the collective character of constituent interactions in the fireball?
\end{itemize}

To answer the above, new experiments are needed to carry out. Of principal importance among them is the search for thermal photons emitted by the 
equilibrium and thermal QGP medium if this medium appears in the nuclei collisions at the RHIC energies. The knowledge of the photon spectrum is 
equal to the direct measurement of the temperature of the irradiating medium \cite{30}. In this case by using the color analogue of the Saha 
formula for ordinary plasma obtained in the work \cite{31}, it is possible to estimate quantitatively the degree of deconfinement.
	
The author of this work thanks S.V. Molodtsov, E.A. Kuraev, R Lednicky, and also D.A. Arkhipkin, Yu.R. Zulkarneeva for valuable remarks concerning 
the text and fruitful discussions of the questions touched upon in the article.

\begin{table}
\caption{Summary of comparative features of the electron plasma and fireball produced in the collisions at RHIC and SPS, CERN.}
\begin{tabular}{ | p{1.0cm} | p{4.0cm} | p{4.0cm} | p{4.0cm} | c |} \hline
 \multicolumn{2}{| p{5.0cm} |}{\textbf{Properties - features of the electron plasma}} & \textbf{Fireball properties – analogues of the electromagnetic plasma}
 & \textbf{Experimental confirmation} & \textbf{References} \\ \hline
 1 & quasineutrality of the total electric charge & neutrality of the total color charge & is present & \cite{14} \\ \hline
 2 & charge separation the plasma volume (dissociation of atoms) & deconfinement of the quarks of nucleon & is absent & - \\ \hline
 3 & collective character of the particle interactions & collective character of quark interactions in the fireball medium & is present & \cite{15},\cite{17} \\ \hline
 4 & Presence of the quasi-particle excitations in the system & presence of quasi-particle excitations in the fireball medium & there are experimental indications
    & \cite{17},\cite{19}-\cite{21} \\ \hline
 5 & field screening of electric charges & screening of the color charge field & there are experimental indications & \cite{25}-\cite{27} \\ \hline
\end{tabular}
\label{tab:tb_01}
\end{table}

\begin{table}
\caption{Summary of the QGP signatures according to its definitions in [\cite{1}-\cite{9}].}
\footnotesize{
\begin{tabular}{ |c| p{4.0cm} | c |c|c|c|c|c|c|c| p{2.5cm} |} \hline
  &                                                        & 1980 & 1983 & 1984/86 & 1984 & 1989 & 1996 & 1999 & 2004 & 2005 \\ \hline
  1 &  Deconfinement of quarks                               &   +  &   +  &    +    &   +  &   +  &   +  &   +  &   +  &   +  \\ \hline
  2 &  Energy necessary for quark deconfinement              &   +  &   +  &    +    &   +  &   -  &   + ($> T_{c}$)  &   +  &   +  &   -  \\ \hline
  3 &  Size of plasma bigger than nucleon size               &   -  &   -  &    +    &   +  &   +  &   -  &   +  &   -  &   +  \\ \hline
  4 &  Independence of quark motion                          &   -  &   +  &    +    &   +  &   +  &   -  &   -  &   -  &   thermalization of constituents \\ \hline
  5 &  Chiral symmetry restoration                           &   -  &   +  &    -    &   -  &   -  &   -  &   +  &   +  &   -  \\ \hline
  6 &  Smallness of quark masses                             &   -  &   +  &    -    &   -  &   -  &   +  &   -  &   -  &   -  \\ \hline
    &  References                         & \cite{1} & \cite{2} & \cite{3} & \cite{4}  & \cite{5}  & \cite{6} & \cite{7} & \cite{8} & \cite{9} \\ \hline
\end{tabular}
}
\textbf{Note: signs ‘‘+/-’’ mean that the indicated feature is or not necessary in the given definition of plasma;
$T_{c}$ – is a temperature of phase transition from the nucleon matter to QGP.}
\label{tab:tb_02}
\end{table}

\end{document}